\pdfoutput=1
\documentclass[12pt,a4paper]{article}

\usepackage{ifthen} 
\newboolean{pdflatex}
\setboolean{pdflatex}{true} 

\newboolean{articletitles}
\setboolean{articletitles}{true} 

\newboolean{uprightparticles}
\setboolean{uprightparticles}{false} 

\def\paperauthors{LHCb collaboration} 
\def\paperasciititle{Search for CP violation in the phase space of D0 -> pi+pi-pi0 decays with LHCb data} 
\def\papertitle{Search for \CP violation in the phase space of $\Dz \to \KS \Kpm \pimp $ decays with the energy test} 
\def\paperkeywords{{High Energy Physics}, {LHCb}} 
\def\papercopyright{\the\year\ CERN for the benefit of the LHCb collaboration} 
\def\paperlicence{CC BY 4.0 licence}
\def\paperlicenceurl{https://creativecommons.org/licenses/by/4.0/}
\def\tsymbol {\ensuremath{T}\xspace}
\def\tvalue {\ensuremath{T}-value\xspace}

\usepackage[top=1in, bottom=1.25in, left=1in, right=1in]{geometry}

\usepackage{microtype}
\usepackage{lineno}  
\usepackage{xspace} 
\usepackage{caption} 

\usepackage{graphicx}  
\usepackage{color}
\usepackage{colortbl}
\graphicspath{{./figs/}} 

\usepackage{amsmath} 
\usepackage{amssymb}
\usepackage{amsfonts}
\usepackage{gensymb}
\usepackage{upgreek} 

\newcommand*\patchAmsMathEnvironmentForLineno[1]{%
\expandafter\let\csname old#1\expandafter\endcsname\csname #1\endcsname
\expandafter\let\csname oldend#1\expandafter\endcsname\csname
end#1\endcsname
 \renewenvironment{#1}%
   {\linenomath\csname old#1\endcsname}%
   {\csname oldend#1\endcsname\endlinenomath}%
}
\newcommand*\patchBothAmsMathEnvironmentsForLineno[1]{%
  \patchAmsMathEnvironmentForLineno{#1}%
  \patchAmsMathEnvironmentForLineno{#1*}%
}
\AtBeginDocument{%
\patchBothAmsMathEnvironmentsForLineno{equation}%
\patchBothAmsMathEnvironmentsForLineno{align}%
\patchBothAmsMathEnvironmentsForLineno{flalign}%
\patchBothAmsMathEnvironmentsForLineno{alignat}%
\patchBothAmsMathEnvironmentsForLineno{gather}%
\patchBothAmsMathEnvironmentsForLineno{multline}%
\patchBothAmsMathEnvironmentsForLineno{eqnarray}%
}

\usepackage{hyperxmp}

\usepackage[pdftex,
            pdfauthor={\paperauthors},
            pdftitle={\paperasciititle},
            pdfkeywords={\paperkeywords},
            pdfcopyright={Copyright (C) \papercopyright},
            pdflicenseurl={\paperlicenceurl}]{hyperref}

\usepackage[colorinlistoftodos,textsize=scriptsize]{todonotes}

\usepackage[bottom,flushmargin,hang,multiple]{footmisc}

\usepackage[all]{hypcap} 

\usepackage{xspace} 
\usepackage{upgreek}

\def\lhcb   {\mbox{LHCb}\xspace}

\def\MagUp {\mbox{\em Mag\kern -0.05em Up}\xspace}

\ifthenelse{\boolean{uprightparticles}}%
{

 \def\Ppi         {\ensuremath{\uppi}\xspace}

 \def\PDelta      {\ensuremath{\Delta}\xspace}                 
 \def\PXi         {\ensuremath{\Xi}\xspace}                 
 \def\PLambda     {\ensuremath{\Lambda}\xspace}                 
 \def\PSigma      {\ensuremath{\Sigma}\xspace}                 
 \def\POmega      {\ensuremath{\Omega}\xspace}                 
 \def\PUpsilon    {\ensuremath{\Upsilon}\xspace}
 \let\oldPi\Pi
 \def\PPi         {\ensuremath{\oldPi}\xspace}

 \def\PB      {\ensuremath{\mathrm{B}}\xspace}                 
                  
 \def\PD      {\ensuremath{\mathrm{D}}\xspace}

 \def\PK      {\ensuremath{\mathrm{K}}\xspace}

 \def\Pb      {\ensuremath{\mathrm{b}}\xspace}                 
 \def\Pc      {\ensuremath{\mathrm{c}}\xspace}

 \def\Pi      {\ensuremath{\mathrm{i}}\xspace}

 \def\Ps      {\ensuremath{\mathrm{s}}\xspace}

 \def\thebaroffset{0.0em}
}
{

 \def\Ppi         {\ensuremath{\pi}\xspace}

 \mathchardef\PDelta="7101
 \mathchardef\PXi="7104
 \mathchardef\PLambda="7103
 \mathchardef\PSigma="7106
 \mathchardef\POmega="710A
 \mathchardef\PUpsilon="7107
 \mathchardef\PPi="7105
                  
 \def\PB      {\ensuremath{B}\xspace}                 
                  
 \def\PD      {\ensuremath{D}\xspace}

 \def\PK      {\ensuremath{K}\xspace}

 \def\Pb      {\ensuremath{b}\xspace}                 
 \def\Pc      {\ensuremath{c}\xspace}

 \def\Pi      {\ensuremath{i}\xspace}

 \def\Ps      {\ensuremath{s}\xspace}

 \def\thebaroffset{0.18em}
}
\newcommand{\offsetoverline}[2][\thebaroffset]{\kern #1\overline{\kern -#1 #2}}%

\makeatletter
\ifcase \@ptsize \relax
  \newcommand{\miniscule}{\@setfontsize\miniscule{4}{5}}
\or
  \newcommand{\miniscule}{\@setfontsize\miniscule{5}{6}}
\or
  \newcommand{\miniscule}{\@setfontsize\miniscule{5}{6}}
\fi
\makeatother

\DeclareRobustCommand{\optbar}[1]{\shortstack{{\miniscule (\rule[.5ex]{1.25em}{.18mm})}
  \\ [-.7ex] $#1$}}

\def\squark    {{\ensuremath{\Ps}}\xspace}

\def\cquark    {{\ensuremath{\Pc}}\xspace}

\def\bquark    {{\ensuremath{\Pb}}\xspace}

\def\pion   {{\ensuremath{\Ppi}}\xspace}

\def\pip    {{\ensuremath{\pion^+}}\xspace}
\def\pim    {{\ensuremath{\pion^-}}\xspace}

\def\pimp   {{\ensuremath{\pion^\mp}}\xspace}

\def\kaon    {{\ensuremath{\PK}}\xspace}

\def\KorKbar {\kern \thebaroffset\optbar{\kern -\thebaroffset \PK}{}\xspace}

\def\Kp      {{\ensuremath{\kaon^+}}\xspace}
\def\Km      {{\ensuremath{\kaon^-}}\xspace}
\def\Kpm     {{\ensuremath{\kaon^\pm}}\xspace}

\def\KS      {{\ensuremath{\kaon^0_{\mathrm{S}}}}\xspace}

\def\Kstarz  {{\ensuremath{\kaon^{*0}}}\xspace}

\def\Kstar   {{\ensuremath{\kaon^*}}\xspace}

\def\Kstarpm {{\ensuremath{\kaon^{*\pm}}}\xspace}
\def\Kstarmp {{\ensuremath{\kaon^{*\mp}}}\xspace}

\def\Dbar    {{\ensuremath{\offsetoverline{\PD}}}\xspace}
\def\D       {{\ensuremath{\PD}}\xspace}

\def\DorDbar {\kern \thebaroffset\optbar{\kern -\thebaroffset \PD}\xspace}
\def\Dz      {{\ensuremath{\D^0}}\xspace}
\def\Dzb     {{\ensuremath{\Dbar{}^0}}\xspace}
\def\Dp      {{\ensuremath{\D^+}}\xspace}
\def\Dm      {{\ensuremath{\D^-}}\xspace}

\def\DpDm    {\ensuremath{\Dp {\kern -0.16em \Dm}}\xspace}

\def\Dstarp  {{\ensuremath{\D^{*+}}}\xspace}

\def\B       {{\ensuremath{\PB}}\xspace}

\def\BorBbar {\kern \thebaroffset\optbar{\kern -\thebaroffset \PB}\xspace}

\def\Bd      {{\ensuremath{\B^0}}\xspace}

\def\BdorBdbar {\kern \thebaroffset\optbar{\kern -\thebaroffset \Bd}\xspace}

\def\Bs      {{\ensuremath{\B^0_\squark}}\xspace}

\def\BsorBsbar {\kern \thebaroffset\optbar{\kern -\thebaroffset \Bs}\xspace}

\def\Y#1S{\ensuremath{\PUpsilon{(#1S)}}\xspace}

\def\LorLbar     {\kern \thebaroffset\optbar{\kern -\thebaroffset \PLambda}\xspace}

\newcommand{\decay}[2]{\ensuremath{#1\!\to #2}\xspace} 

\def\to                 {\ensuremath{\rightarrow}\xspace}

\def\CP                {{\ensuremath{C\!P}}\xspace}

\def\AT#1     {\ensuremath{A_{\mathrm{T}}^{#1}}\xspace}           

\def\C#1      {\ensuremath{\mathcal{C}_{#1}}\xspace}                       
\def\Cp#1     {\ensuremath{\mathcal{C}_{#1}^{'}}\xspace}                    
\def\Ceff#1   {\ensuremath{\mathcal{C}_{#1}^{\mathrm{(eff)}}}\xspace}        
\def\Cpeff#1  {\ensuremath{\mathcal{C}_{#1}^{'\mathrm{(eff)}}}\xspace}       
\def\Ope#1    {\ensuremath{\mathcal{O}_{#1}}\xspace}                       
\def\Opep#1   {\ensuremath{\mathcal{O}_{#1}^{'}}\xspace}                    

\newcommand{\nospaceunit}[1]{\ensuremath{\text{#1}}}       
\newcommand{\aunit}[1]{\ensuremath{\text{\,#1}}}       

\newcommand{\tev}{\aunit{Te\kern -0.1em V}\xspace}
\newcommand{\gev}{\aunit{Ge\kern -0.1em V}\xspace}
\newcommand{\mev}{\aunit{Me\kern -0.1em V}\xspace}
\newcommand{\kev}{\aunit{ke\kern -0.1em V}\xspace}
\newcommand{\ev}{\aunit{e\kern -0.1em V}\xspace}
 
\newcommand{\mevc}{\ensuremath{\aunit{Me\kern -0.1em V\!/}c}\xspace}
\newcommand{\gevc}{\ensuremath{\aunit{Ge\kern -0.1em V\!/}c}\xspace}
\newcommand{\mevcc}{\ensuremath{\aunit{Me\kern -0.1em V\!/}c^2}\xspace}
\newcommand{\gevcc}{\ensuremath{\aunit{Ge\kern -0.1em V\!/}c^2}\xspace}

\def\mum  {\ensuremath{\,\upmu\nospaceunit{m}}\xspace}

\def\fb   {\ensuremath{\aunit{fb}}\xspace}
\def\invfb   {\ensuremath{\fb^{-1}}\xspace}

\def\gsim{{~\raise.15em\hbox{$>$}\kern-.85em
          \lower.35em\hbox{$\sim$}~}\xspace}
\def\lsim{{~\raise.15em\hbox{$<$}\kern-.85em
          \lower.35em\hbox{$\sim$}~}\xspace}

\def\pt         {\ensuremath{p_{\mathrm{T}}}\xspace}

\def\ptot       {\ensuremath{p}\xspace}

\def\tell1  {TELL1\xspace}
\def\ukl1   {UKL1\xspace}

\newcommand{\lhcborcid}[1]{\href{https://orcid.org/#1}{\hspace*{0.1em}\raisebox{-0.45ex}{\includegraphics[width=1em]{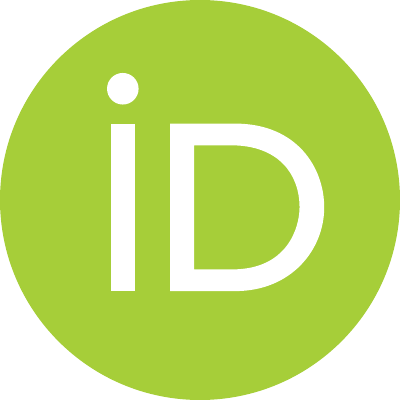}}}}

\usepackage{cite} 
\usepackage{mciteplus}

\usepackage{longtable} 
\usepackage{float}
\begin{document}

\renewcommand{\thefootnote}{\fnsymbol{footnote}}
\setcounter{footnote}{1}


\begin{titlepage}
\pagenumbering{roman}

\vspace*{-1.5cm}
\centerline{\large EUROPEAN ORGANIZATION FOR NUCLEAR RESEARCH (CERN)}
\vspace*{1.5cm}
\noindent
\begin{tabular*}{\linewidth}{lc@{\extracolsep{\fill}}r@{\extracolsep{0pt}}}
\ifthenelse{\boolean{pdflatex}}
{\vspace*{-1.5cm}\mbox{\!\!\!\includegraphics[width=.14\textwidth]{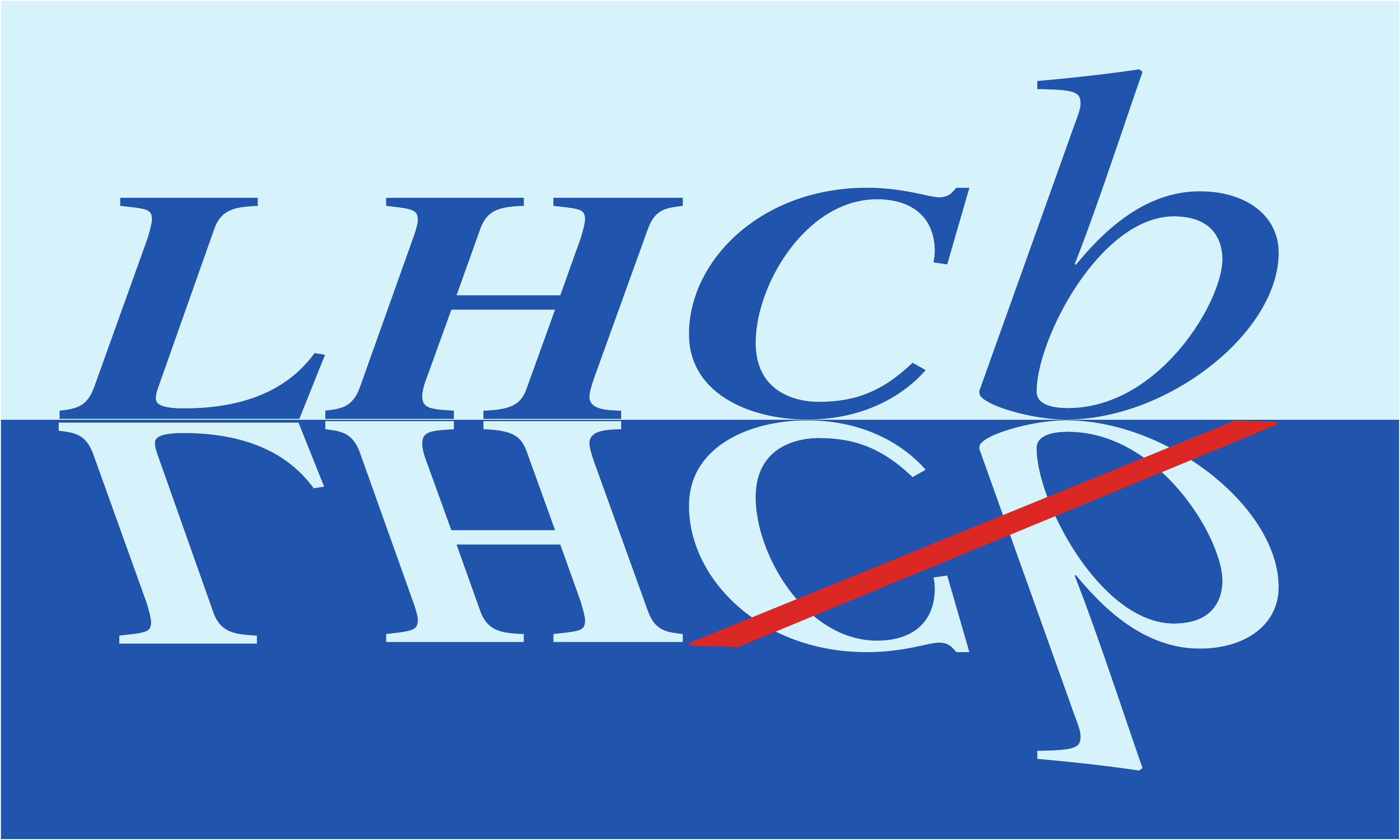}} & &}%
{\vspace*{-1.2cm}\mbox{\!\!\!\includegraphics[width=.12\textwidth]{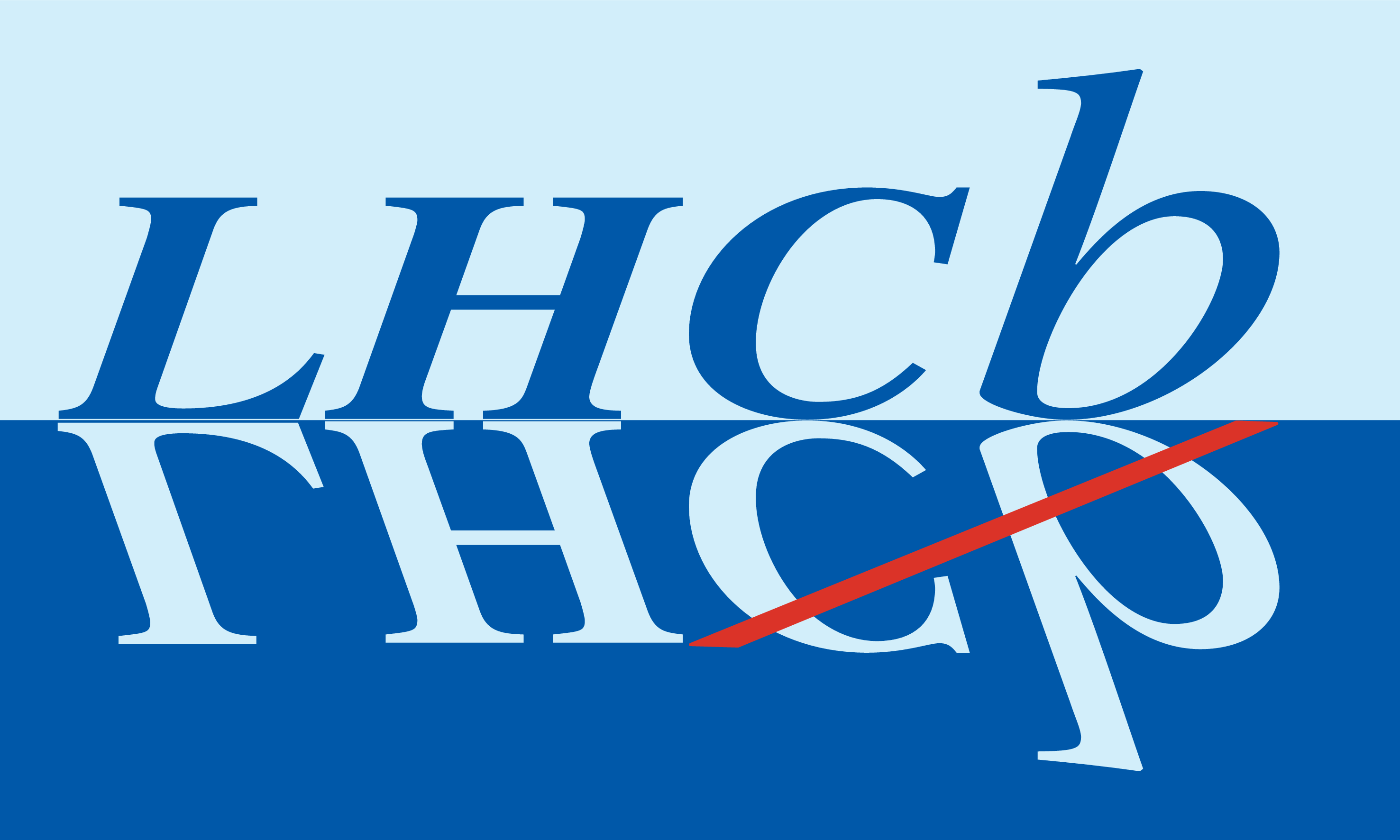}} & &}%
\\
 & & CERN-EP-2023-231 \\  
 & & LHCb-PAPER-2023-019 \\  
 & & March 28, 2024 \\ 
 & & \\
\end{tabular*}

\vspace*{3.0cm}

{\normalfont\bfseries\boldmath\huge
\begin{center}
  \papertitle 
\end{center}
}

\vspace*{1.0cm}

\begin{center}
\paperauthors\footnote{Authors are listed at the end of this paper.}

\vspace*{0.5cm}

This paper is dedicated to the memory of our friend and colleague \\ Alvaro Gomes.

\end{center}

\vspace{\fill}

\begin{abstract}
  \noindent
A search for \CP violation in $\Dz \to \KS \Kp \pim$ and $\Dz \to \KS \Km \pip$ decays is reported. The search is performed using an unbinned model-independent method known as the energy test that probes local \CP violation in the phase space of the decays. The data analysed correspond to an integrated luminosity of $5.4$~\invfb collected in proton-proton collisions by the \lhcb experiment at a centre-of-mass energy of $\sqrt{s}=13 \tev$, amounting to approximately 950 thousand and 620 thousand signal candidates for the $\Dz \to \KS \Km \pip$ and $\Dz \to \KS \Kp \pim$ modes, respectively. The method is validated using $\Dz \to \Km \pip \pim \pip$ and $\Dz \to \KS \pip \pim$ decays, where \CP-violating effects are expected to be negligible, and using background-enhanced regions of the signal decays. The results are consistent with \CP symmetry in both the $\Dz \to \KS \Km \pip$ and the $\Dz \to \KS \Kp \pim$ decays, with $p$-values for the hypothesis of no \CP violation of 70\% and 66\%, respectively.

\end{abstract}

\vspace*{1.0cm}

\begin{center}
  Published in JHEP 03 (2024) 107 
\end{center}

\vspace{\fill}

{\footnotesize 
\centerline{\copyright~\papercopyright. \href{\paperlicenceurl}{\paperlicence}.}}
\vspace*{2mm}

\end{titlepage}


\newpage
\setcounter{page}{2}
\mbox{~}
%
%
%
%


\renewcommand{\thefootnote}{\arabic{footnote}}
\setcounter{footnote}{0}

\cleardoublepage


\pagestyle{plain} 
\setcounter{page}{1}
\pagenumbering{arabic}


\newpage

\section{Introduction}
In the Standard Model (SM) of particle physics, \CP violation and matter-antimatter asymmetries are introduced in the quark sector through a complex phase in the Cabibbo-Kobayashi-Maskawa (CKM) matrix~\cite{Cabibbo:1963yz,Kobayashi:1973fv}. However, the level of \CP violation observed in the SM is not sufficient to explain the measured baryon asymmetry in the universe, motivating the search for new physics that can introduce additional sources of \CP violation~\cite{CP1,CP2}. Since \CP-violating effects in charm decays are expected to be small within the SM, such decays offer an excellent laboratory to search for these new sources. 

While \CP-violating effects have been extensively measured and tested in the beauty and strange sectors, \CP violation in charm has only recently been observed, with the measurement of the difference between asymmetries in two singly Cabibbo-suppressed (SCS) modes~\cite{LHCb-PAPER-2019-006}. It is not yet clear whether this result is consistent with the SM, or if extensions to the SM are required to explain the measurement~\cite{Chala:2019fdb,Li:2019hho, Grossman:2019xcj,Soni:2019xko, Cheng:2019ggx}. Further analysis of \CP violation in the charm sector is therefore essential. Multibody charm decays offer an excellent environment for such studies. Large, localised \CP-violating effects can arise in multibody decays, because of the interplay of different intermediate resonances and corresponding variations of the strong phase across the relevant phase space.

This paper presents a search for \CP violation in \mbox{$\Dz \to \KS \Kpm \pimp$} decays.\footnote{Throughout this paper charge conjugation is implied.} These decays are also dominated by SCS amplitudes, such that they offer excellent prospects to further understand the observation of \CP violation in the charm sector. These decays also include significant contributions from \mbox{$\Dz \to \Kstarmp \Kpm$} decays. In addition, the decay modes under study contain contributions from \mbox{$\Dz\to\KS\Kstarz$} decays, where additional \CP-violating effects in the charm sector can arise~\cite{paper_kkstar}. 
The \mbox{$\Dz \to \KS \Kpm \pimp$} decay modes have been previously studied at LHCb~\cite{LHCb-PAPER-2015-026}, in an amplitude analysis using a dataset corresponding to 3\invfb of integrated luminosity collected in the first run of Large Hadron Collider (LHC) operations at centre-of-mass energies $\sqrt{s} = 7$~and~$8\tev$. The analysis presented here complements that study with a search for local \CP-violating effects using an unbinned, model-independent method, called the energy test~\cite{zech1,zech2,Williams:2011cd,Parkes:2016yie}. This method has been applied in previous analyses of charm and beauty hadron decays at LHCb~\cite{LHCb-PAPER-2014-054,LHCb-PAPER-2016-044,LHCb-PAPER-2019-028,LHCb-PAPER-2023-005}. The new study reported here makes use of 5.4\invfb of data collected in the second period of LHC operations at $\sqrt{s} = 13\tev$, with signal yields about seven times larger than the previous LHCb amplitude analysis of these decay modes~\cite{LHCb-PAPER-2015-026}.

\section{Analysis overview}
\label{sec:overview}
In this study, samples of \Dz mesons decaying to the final state \KS \Kpm \pimp are analysed using the energy test. The method tests a hypothesis of no \CP violation. The final result is a single $p$-value that describes the consistency with this hypothesis. 

Candidate \Dz decays are identified by selecting \Dstarp candidates that undergo the strong decay \decay{\Dstarp}{\Dz\pip}. The flavour of the \Dz or \Dzb candidate at production is directly determined by the charge of the $\pip$ (denoted the soft pion) in this decay. 
The \Dz meson then decays to the final state \KS \Kp \pim or \KS \Km \pip. 
The decay is referred to as ``same sign'' (SS) when the charge of the pion produced in the $\Dz$ meson decay matches that of the soft pion in the \mbox{$\Dstarp \to \Dz \pip$} decay. If the pion from the \Dz meson has the opposite charge to the soft pion, the final state is referred to as the ``opposite sign'' (OS) decay. There are therefore four different decay modes in total: \mbox{$\Dz \to \KS \Kp \pim$} (OS), \mbox{$\Dz \to \KS \Km \pip$} (SS), \mbox{$\Dzb \to \KS \Kp \pim$} (SS) and \mbox{$\Dzb \to \KS \Km \pip$} (OS).
The branching fraction of the SS decays are approximately a factor 1.5 larger than that of the OS decays~\cite{PDG2022}. The \KS meson is selected via its decay to the \pip\pim final state. To minimise detector effects that could mimic potential \CP violation in these decay modes, two other decay modes that are dominated by Cabibbo-favoured amplitudes are employed as control channels: \mbox{$\Dz \to \Km \pip \pim \pip$} and \mbox{$\Dz \to \KS \pip \pim$}. Both control channels are topologically similar to the signal modes and no significant \CP-violating effects are expected in these transitions~\cite{Grossman:2006jg}. 

The phase space of the \mbox{$\Dz \to \KS \Kpm \pimp$} decay is parameterised in terms of the invariant mass squared of pairs of the decay products: $s_{12} = m^2(\KS \Kpm)$, $s_{13} = m^2(\KS \pimp)$, and $s_{23} = m^2(\Kpm \pimp)$. While only two quantities are necessary to fully parameterise a three-body decay, the use of all three two-body combinations to describe the phase space in this study ensures that each quantity is treated on an equal footing. 

\section{Energy test method}
\label{sec:energytest}
The energy test is an unbinned, model-independent method that tests if two multi-dimensional datasets are consistent with the same underlying distribution~\cite{zech1,zech2,Williams:2011cd,Parkes:2016yie}.
The comparison of \Dz and \Dzb decays using this method therefore provides a powerful search for local \CP violation. The test quantifies differences within the phase space of the two samples, through the calculation of a test statistic, the \tvalue, determined as
\begin{equation}
    \tsymbol \equiv  \dfrac{1}{2n(n-1)}\sum_{i,j\neq i}^{n}\psi_{ij}
+ \dfrac{1}{2\overline{n}(\overline{n}-1)}\sum_{i,j\neq i}^{\overline{n}}\psi_{ij}
- \dfrac{1}{n\overline{n}} \sum_{i,j}^{n,\overline{n}}\psi_{ij}.
\label{eq:t}
\end{equation}
The first term is a sum over all pairs of ($n$) \Dz candidates, the second term is a sum over all pairs of ($\overline{n}$) \Dzb candidates, while the final term considers all \Dz-\Dzb pairs. The function $\psi_{ij}$ provides a weight to each pair of candidates, and is taken to be $\psi_{ij}=e^{-d_{ij}^2/2\delta^2}$, where $d_{ij}^2 = (s_{12,i} - s_{12,j})^2 + (s_{13,i} - s_{13,j})^2 + (s_{23,i} - s_{23,j})^2$ is the square of the Euclidean distance between the two candidates under consideration, using the parameterisation of the phase space discussed in Sec.~\ref{sec:overview}. The tuneable parameter $\delta$ sets the distance scale probed in the study, and is optimised to maximise the sensitivity to \CP violation, as discussed in Section~\ref{sec:optimisation}. The normalisation of each term in Eq.~\ref{eq:t} ensures that the test is only sensitive to local sample differences, and is, by construction, insensitive to global asymmetries. The value of $T$ is expected to be close to zero under the null hypothesis of no sample differences, but to be large and positive in the presence of significant \CP violation.

The significance of the calculated \tvalue is determined by comparing the value found to the distribution under the null hypothesis. This distribution depends on the specifics of the sample under study, so is determined using a permutation method, repeatedly running the test using the recorded data but with the flavour of the \Dz and \Dzb mesons randomly assigned. This stage can be performed with a subset of the signal sample, significantly increasing the speed with which this distribution can be determined, following advances in Refs.~\cite{Barter:2018xbc,Zech:2018bpy,Gillam:2018ufy}.
The fraction of these permutation samples with a \tvalue greater than that observed in the (unpermuted) datasets is the $p$-value associated with the study.

\section{LHCb experiment}
The \lhcb detector~\cite{LHCb-DP-2008-001,LHCb-DP-2014-002} is a single-arm forward spectrometer covering the \mbox{pseudorapidity} range $2<\eta <5$, designed for the study of particles containing \bquark or \cquark
quarks. The detector includes a high-precision tracking system consisting of a silicon-strip vertex detector surrounding the proton-proton interaction region~\cite{LHCb-DP-2014-001}, a large-area silicon-strip detector located upstream of a dipole magnet with a bending power of about $4{\mathrm{\,Tm}}$, and three stations of silicon-strip detectors and straw drift tubes~\cite{LHCb-DP-2017-001} placed downstream of the magnet. The tracking system provides a measurement of the momentum, \ptot, of charged particles with a relative uncertainty that varies from 0.5\% at low momentum to 1.0\% at 200\gevc. The minimum distance of a track to a primary proton-proton collision vertex (PV), the impact parameter (IP), is measured with a resolution of $(15+29/\pt)\mum$, where \pt is the component of the momentum transverse to the beam, in\,\gevc. Different types of charged hadrons are distinguished using information from two ring-imaging Cherenkov detectors~\cite{LHCb-DP-2012-003}. Photons, electrons and hadrons are identified by a calorimeter system consisting of scintillating-pad and preshower detectors, an electromagnetic and a hadronic calorimeter. Muons are identified by a
system composed of alternating layers of iron and multiwire proportional chambers~\cite{LHCb-DP-2012-002}. 

The magnetic field deflects oppositely charged particles in opposite
directions, which can lead to detection asymmetries. Periodically
reversing the magnetic field polarity throughout the data-taking almost cancels
the effect. 

The online event selection is performed by a trigger, which consists of a hardware stage based on information from the calorimeter and muon systems followed by two software stages. In between the software stages, an alignment and calibration of the detector is performed in near real-time, with the results then used in the trigger~\cite{LHCb-PROC-2015-011}. The same alignment and calibration information is propagated to the offline reconstruction, ensuring consistent and high-quality particle identification (PID) information between the trigger and offline software. The identical performance of the online and offline reconstruction offers the opportunity to perform physics analyses directly using candidates reconstructed in the trigger \cite{LHCb-DP-2012-004,LHCb-DP-2016-001} which the present analysis exploits. The storage of only the information related to triggered candidates enables a reduction in the event size by an order of magnitude. Candidates are only considered if other particles in the same event are responsible for the hardware trigger decision, to avoid potential biases introduced when signal candidates are responsible for the event being selected.

\section{Signal selection}
\label{sec:selection}
The selection of candidates begins with the trigger stage, where the selection is based on the transverse energy measured by the calorimeters, accepting only particles with a significant energy deposit.
In the first software trigger stage at least one of the particles in the \Dz decay is required to have high transverse momentum and to be inconsistent with originating from a PV~\cite{LHCb-PUB-2011-003}. A dedicated selection of signal decays is then made at the second software trigger level. From this stage onward \decay{\KS}{\pip\pim} decays are considered in two different categories: the first involving \KS mesons that decay early enough for the pions to be reconstructed in the vertex detector; and the second containing \KS mesons that decay later such that track segments of the pions cannot be formed in the vertex detector. These categories are referred to as \textit{long} and \textit{downstream}, respectively. To avoid contamination from \mbox{$\Dz \to K^- \pi^+ \pi^- \pi^+$} decays, the \KS vertex is required to be significantly displaced from the \Dz meson decay vertex. This also reduces the background due to other pions produced in the proton-proton collision that are incorrectly reconstructed as a \KS meson. The pions from the \KS decay are required to have $p>3000\mevc$ and $\pt>175\mevc$ if the decay is in the downstream category. The other hadrons produced in the \Dz and \Dstarp decays are required to have $p>1000\mevc$ and $\pt>200\mevc$ irrespective of the \KS decay category, and the \Dz meson is required to have $\pt>1800\mevc$. Particle identification criteria are placed on the pion and kaon directly produced in the \Dz decay, in order to reduce background. 

To improve the purity of the data sample and reduce any residual detector asymmetries, additional criteria are applied offline. 
The \Dz meson is required to be consistent with originating at the PV, and to decay significantly displaced from this point.  As low-momentum charged particles might be swept away from the beam pipe or bent back into it by the magnet, or into and out of the coverage of the LHCb detector, some areas are reached by particles of only one charge for each magnet polarity. This effect can potentially produce local asymmetries in the soft pion phase space. To avoid this the areas of maximal asymmetry are removed from the sample, using criteria inspired by Ref.~\cite{LHCb-PAPER-2013-041}. 
To exclude candidates originating from hadronic interactions with the detector material, a requirement on the radial distance from the beamline is applied on the \Dstarp and \Dz origin and decay vertices. Furthermore, only proton-proton interactions within the main interaction region along the beamline are considered.  
Particle identification criteria are applied to the particles produced in the \Dz and \KS meson decays. Looser particle identification requirements are placed on the soft pions given the lower momenta of these particles.

The signal sample is then defined by selecting candidates with invariant masses within intervals around the mean values of the particles considered. The relevant final-state particles are required to have an invariant mass within $11.6\mevcc$ of the known \KS meson mass~\cite{PDG2022}, within $22.1\mevcc$ of the known \Dz meson mass~\cite{PDG2022}, and to have $\Delta M \equiv m(\Dz\pi^+)-m(\Dz)$ within $0.7\mevcc$ of the known mass difference between the \Dstarp and \Dz meson. 
These intervals correspond to three standard deviations of the detector resolution for the candidate masses in signal decays. A kinematic fit is also performed to better resolve the momenta of the final state particles. The pion produced in the \Dstarp meson decay is constrained to originate from the measured PV. In addition, the \Dz and \KS masses are constrained to their known values~\cite{PDG2022}. After the aforementioned criteria are applied, some events contain multiple candidates. In these events one candidate is selected randomly. This results in a net loss of $\sim 4\%$ for both the SS and OS samples. In total, about 950 thousand SS candidates and about 620 thousand OS candidates are selected; for the SS sample $n = 468\,223$ and $\overline{n} = 480\,059$  whereas for the OS sample $n=310\,701$ and $\overline{n} =310\,480 $.

\begin{figure}[!thb]
    \centering
    \includegraphics[width=0.49\linewidth]{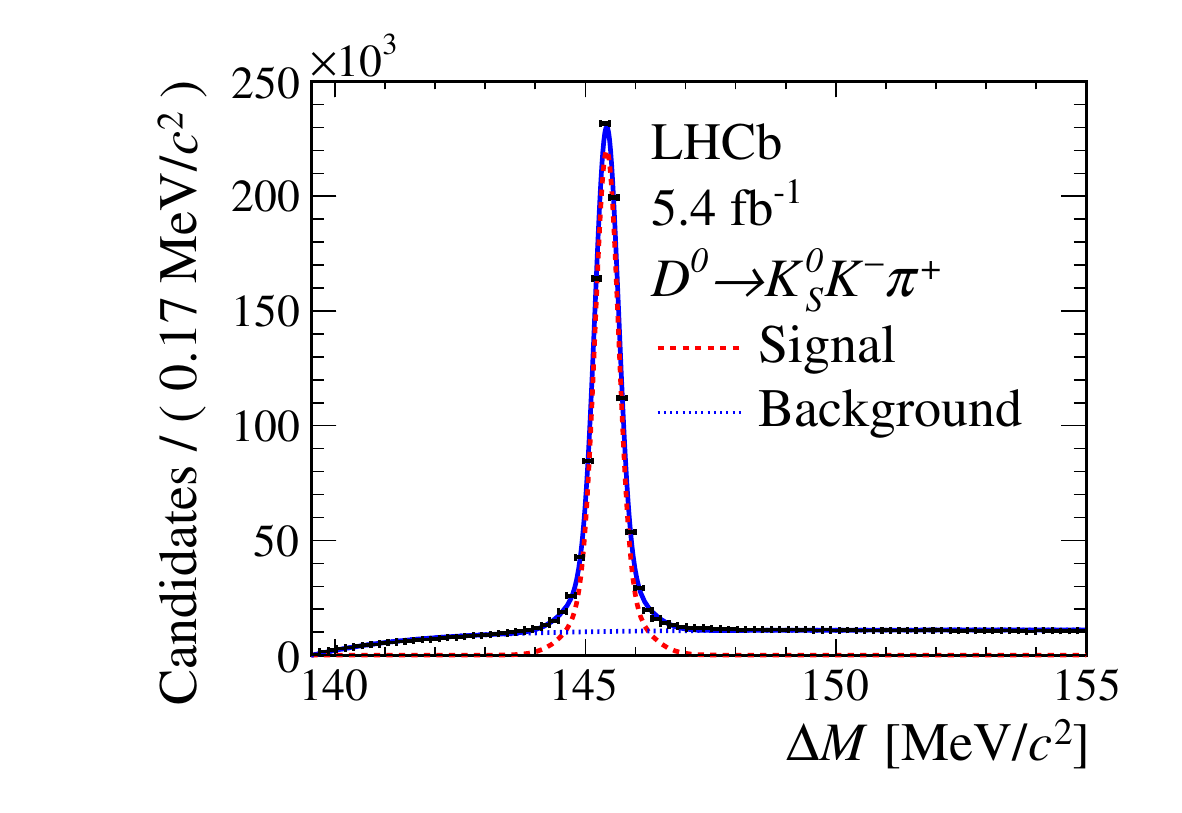}
    \includegraphics[width=0.49\linewidth]{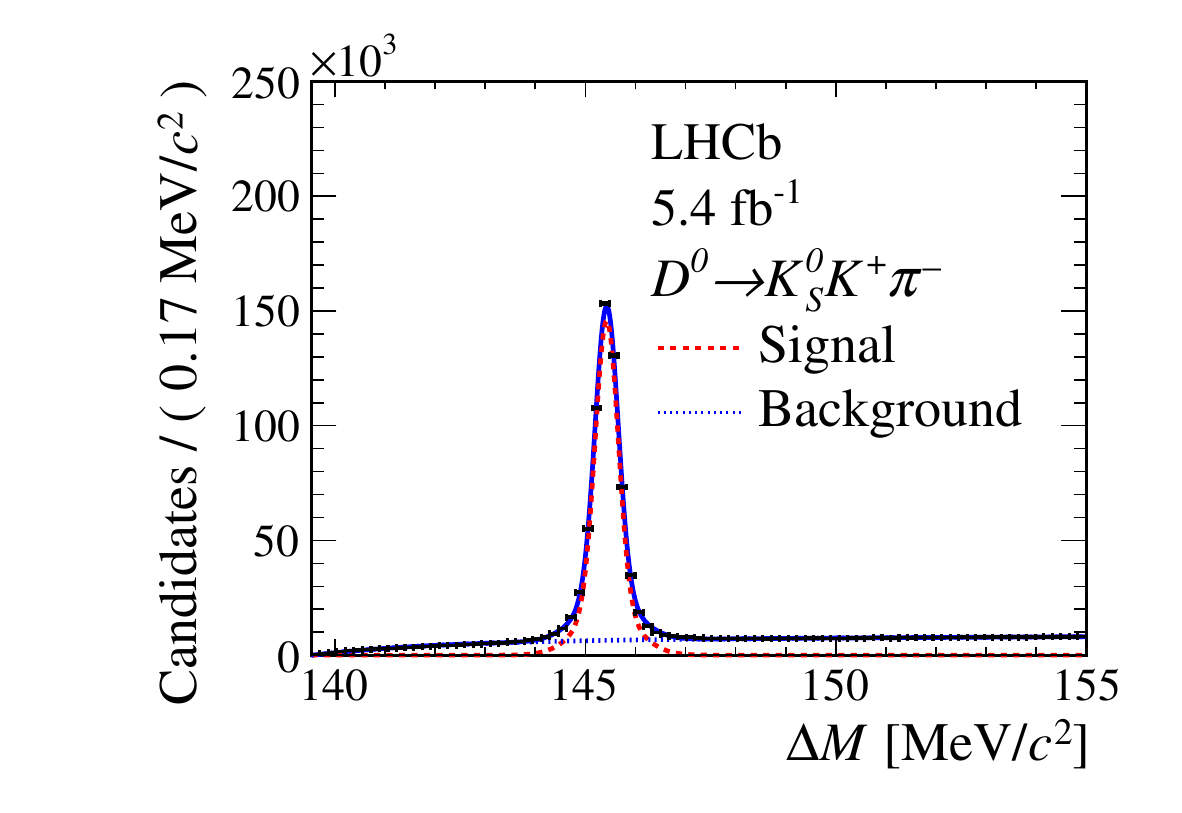}
    \caption{Distributions of $\Delta M$ for \mbox{$D^{0} \to K_{S}^{0} K^{\pm} \pi^{\mp}$} decays for (left) SS  and (right) OS  decays with fit results overlaid. For each decay charge conjugation is always implied.}
    \label{fig:signal_yields_DTF}
\end{figure}

\begin{figure}[!thb]
    \centering
        \includegraphics[width=0.49\linewidth]{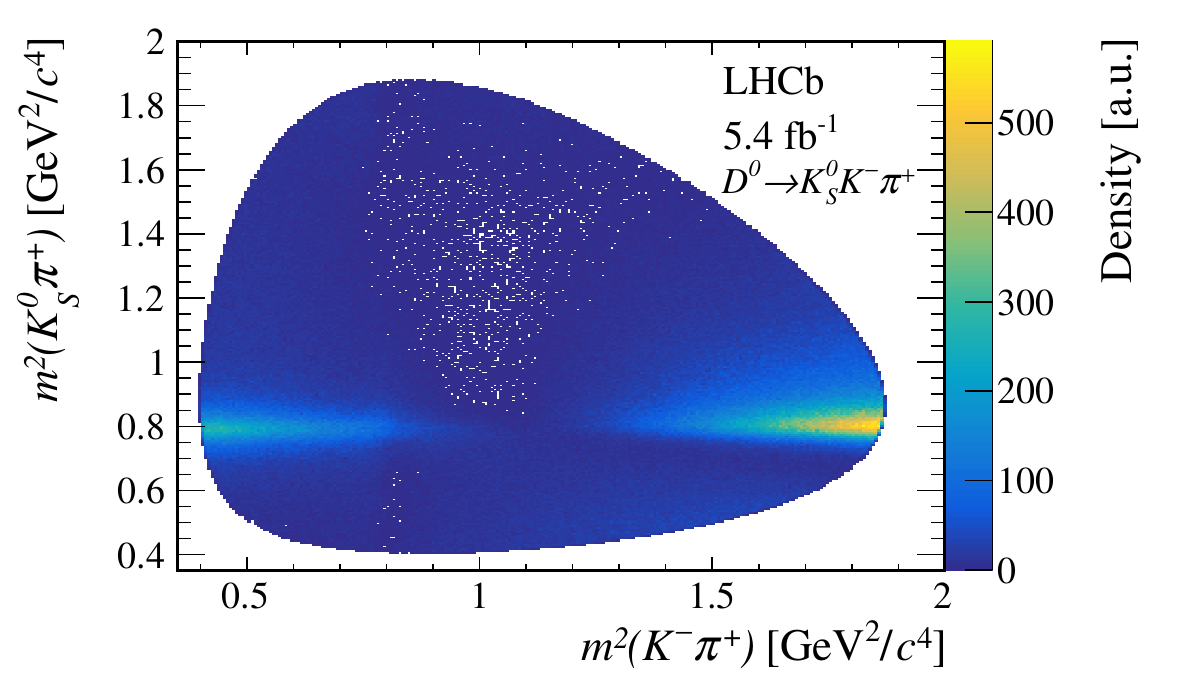}
        \includegraphics[width=0.49\linewidth]{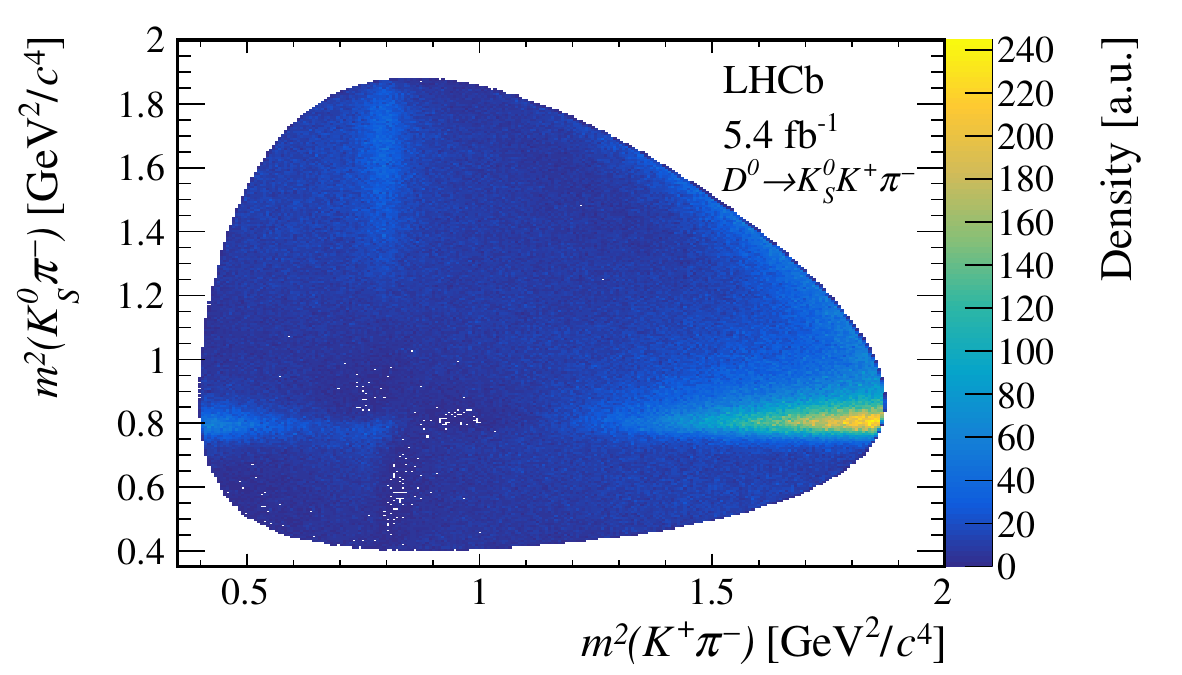} 
    \caption{Dalitz plots of the selected signal samples for the (left) SS and (right) OS  decays. The peaking structure seen  as a horizontal band in both plots corresponds to the \Kstarpm resonance. The \Kstarz resonance is also visible as a vertical band in the OS decays phase space distribution. For each decay charge conjugation is always implied.}
    
    \label{fig:final_data_dalitz}
\end{figure}

The purity of the samples under study are estimated with a binned maximum-likelihood fit of the $\Delta M$ distribution. The fit uses a specific probability density function (PDF) to model the background component following Ref.~\cite{LHCb-PAPER-2012-038}, and uses the sum of two Gaussian functions to model the signal component. These two Gaussian functions are constrained to the same mean value while the widths are free parameters. The distributions of $\Delta M$ for the SS and OS samples are shown in Fig.~\ref{fig:signal_yields_DTF}, along with the results of the fit. The purity within the defined signal mass window is about $91\%$ for both the SS sample and the OS sample. 
Figure~\ref{fig:final_data_dalitz} shows the Dalitz plots of the selected signal samples.

\section{Optimisation and sensitivity studies}
\label{sec:optimisation}

The energy test is associated with a single free parameter, $\delta$, which sets the distance-scale associated with the test, as discussed in Sec.~\ref{sec:energytest}. Different choices can be made for this parameter, with these choices then providing different sensitivity to local \CP violation. To maximise the power of the test applied here to potential \CP-violating effects, simulated pseudoexperiments are performed. These pseudoexperiments make use of the $\Dz \to \KS \Kpm \pimp$ amplitude model determined using data collected at the LHCb experiment in the first period of LHC operations~\cite{LHCb-PAPER-2015-026}, and are generated using the GOOFIT package~\cite{GOOFIT}. The dataset used in the analysis presented here does not overlap with the dataset that was used to determine this amplitude model. 

Different \CP asymmetries are introduced in these studies by varying the amplitudes associated with specific contributing resonances, with the magnitudes varied between 1\% and 12\%, and the phases varied between 1$^\circ$ and 14$^\circ$. The generated pseudoexperiment samples have a comparable sample size to that of the LHCb dataset analysed here, and also include a variation of efficiency across the phase space of the decay, based on that expected in data. The energy test is run on these simulated datasets, and the $p$-value for the null hypothesis of \CP symmetry is recorded for different values of $\delta$. The value of $\delta$ which gives the smallest $p$-value is then considered the optimal choice for the variation under study. For most resonances tested the minimum $p$-value is achieved for $\delta\sim 0.2\gev^2/c^4$, so this value is selected as the baseline value for analysing the real data, to ensure the most sensitive test is performed.
This approach also enables the sensitivity of the OS and SS analyses carried out here to be confirmed and estimated. 
For example, in both the SS and OS modes, a $p$-value of $\mathcal{O}(10^{-4})$ is expected if a 2\% \CP asymmetry is present in the magnitude of the amplitude associated with the $\Kstar(892)^\pm$ resonances, while a $p$-value of $\mathcal{O}(10^{-3})$ is expected for a $2^\circ$ phase difference between the \CP-conjugate decays.

\section{Control channels}
The method described in the previous sections is tested using control channels to confirm that any observed sample differences are not caused by instrumentation effects, or by \CP violation in the \KS meson decay. To confirm that instrumentation effects do not mimic \CP violation, the $\Dz \to \Km \pip \pim \pip$ decay is studied. This decay is dominated by Cabibbo-favoured contributions, ensuring a large sample size, and implying that the expected level
of CP-violation is negligible~\cite{Grossman:2006jg}. Likewise, $\Dz \to \KS \pip \pim$ decays are studied to confirm negligible effects from asymmetries in the \KS meson decays. In both cases the \Dz meson flavour is identified through the same \Dstarp decay.
Both control channels are selected using criteria closely reproducing the ones used for the signal channel. 

\begin{figure}[!tbh]
    \centering
    \includegraphics[width=0.45\linewidth]{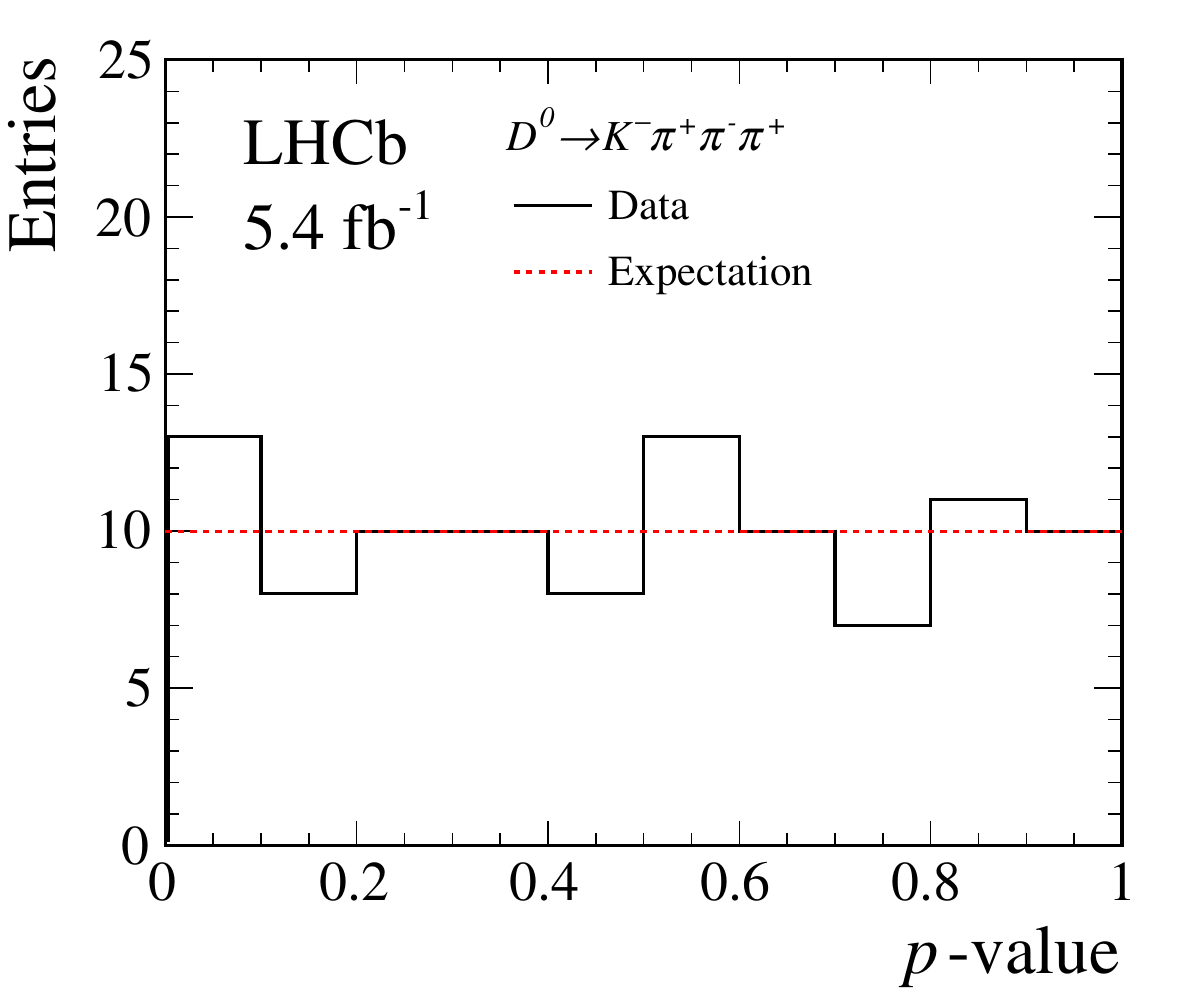}
    \includegraphics[width=0.45\linewidth]{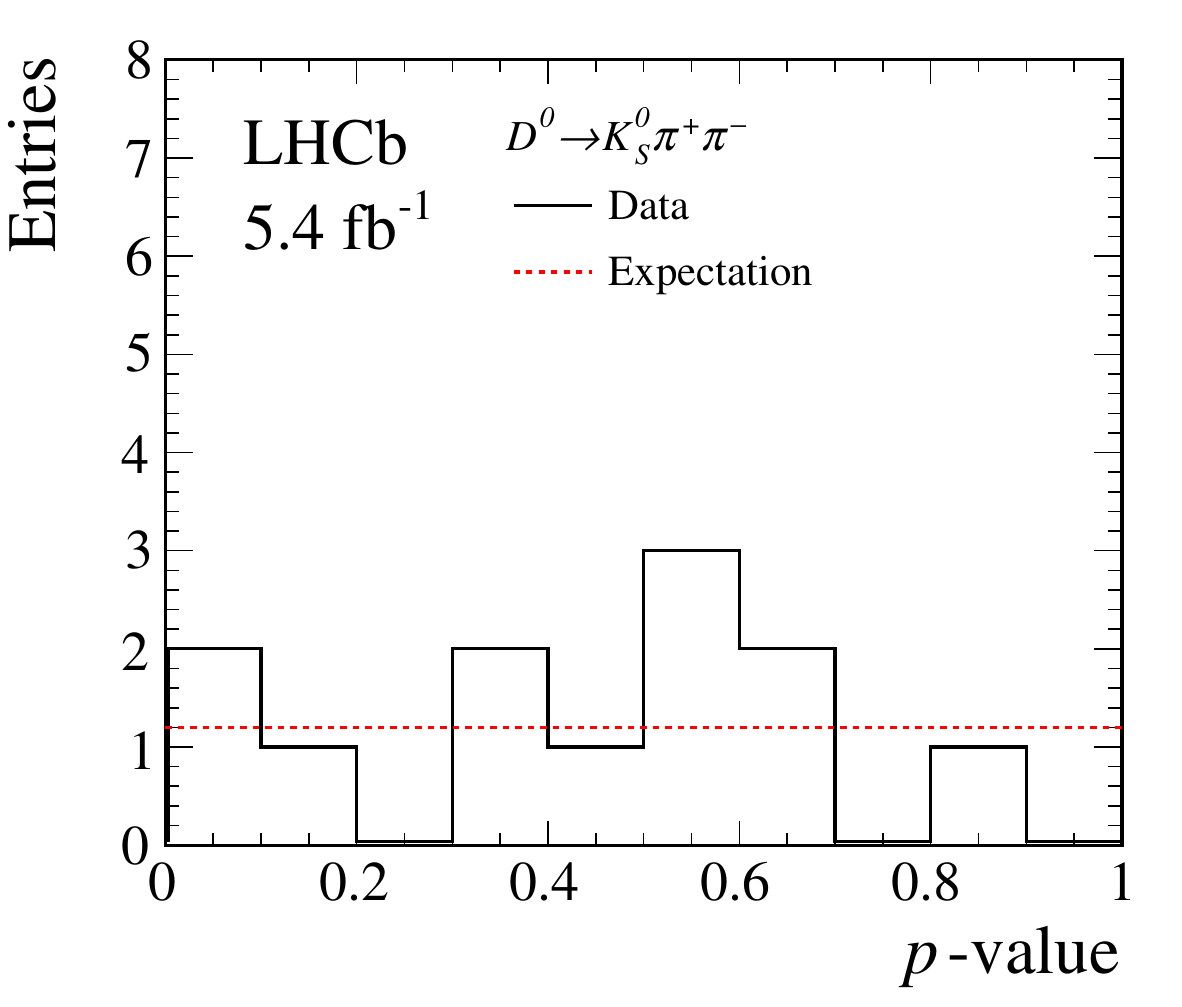}
    \caption{Distribution of $p$-values for the control channels obtained from (left) \mbox{$\Dz \to \Km \pip \pim \pip$} decays and (right) $\Dz \to \KS \pip \pim$ decays. Each $p$-value is obtained by comparing subsamples with roughly the same yield as that observed in the signal channel.}
    \label{fig:control_channel_delta02}
\end{figure}

The yield in the $\Dz \to \Km \pip \pim \pip$ decay mode is roughly one hundred times larger than in the signal decay modes, while the $\Dz \to \KS \pip \pim$ yield is about twelve times the signal yield.  Therefore subsamples are randomly produced in the control channels, each containing the same number of candidates as the signal yield in the SS channel. Each of these subsamples is then tested individually using the energy test. The control channel tests therefore result in a distribution of $p$-values. As the control samples are expected to be \CP symmetric, the resulting $p$-values should be uniformly distributed, while local biases would lead to a non-uniform distribution.
The results of these tests are shown in Fig.~\ref{fig:control_channel_delta02}. Potential deviations from a flat distribution are probed using $\chi^2$-tests, which return $p$-values of 94\% and 43\% for the $\Dz \to \Km \pip \pim \pip$ decay mode and $\Dz \to \KS \pip \pim$ decay mode, respectively. No significant deviation from a flat distribution is therefore observed, and potential biases are considered negligible.

Further checks have been carried out. First, the previous test is repeated using subsamples containing double the number of candidates in the signal channel. The results are again consistent with negligible bias. In addition, the control channels are confirmed to exhibit no evidence for potential biases when the dataset is split into different data-taking periods, both by the year and by the direction of the magnetic field. Further, decays of the \KS meson are split according to the location of the \KS decay in the detector. Again, no evidence for bias is seen. Finally, the \mbox{$\Dz \to \Km \pip \pim \pip$} decay is randomly assigned a flavour, and then the known $K$-$\pi$ instrumentation asymmetry~\cite{LHCb-PUB-2018-004} is injected. The same exercise of splitting into appropriately sized subsamples is performed, and the energy test study is repeated. The recorded $p$-value distribution is consistent with no bias.

\section{Backgrounds}
Despite a signal purity that exceeds $90\%$ for both the SS and OS decay modes, the presence of asymmetric background contributions can also induce a large \tvalue and mimic \CP violation. Similar tests to those made using the control channels are therefore made using background-enhanced samples to confirm that the background does not cause such effects.

\begin{figure}[t]
    \centering
    \includegraphics[width=0.45\linewidth]{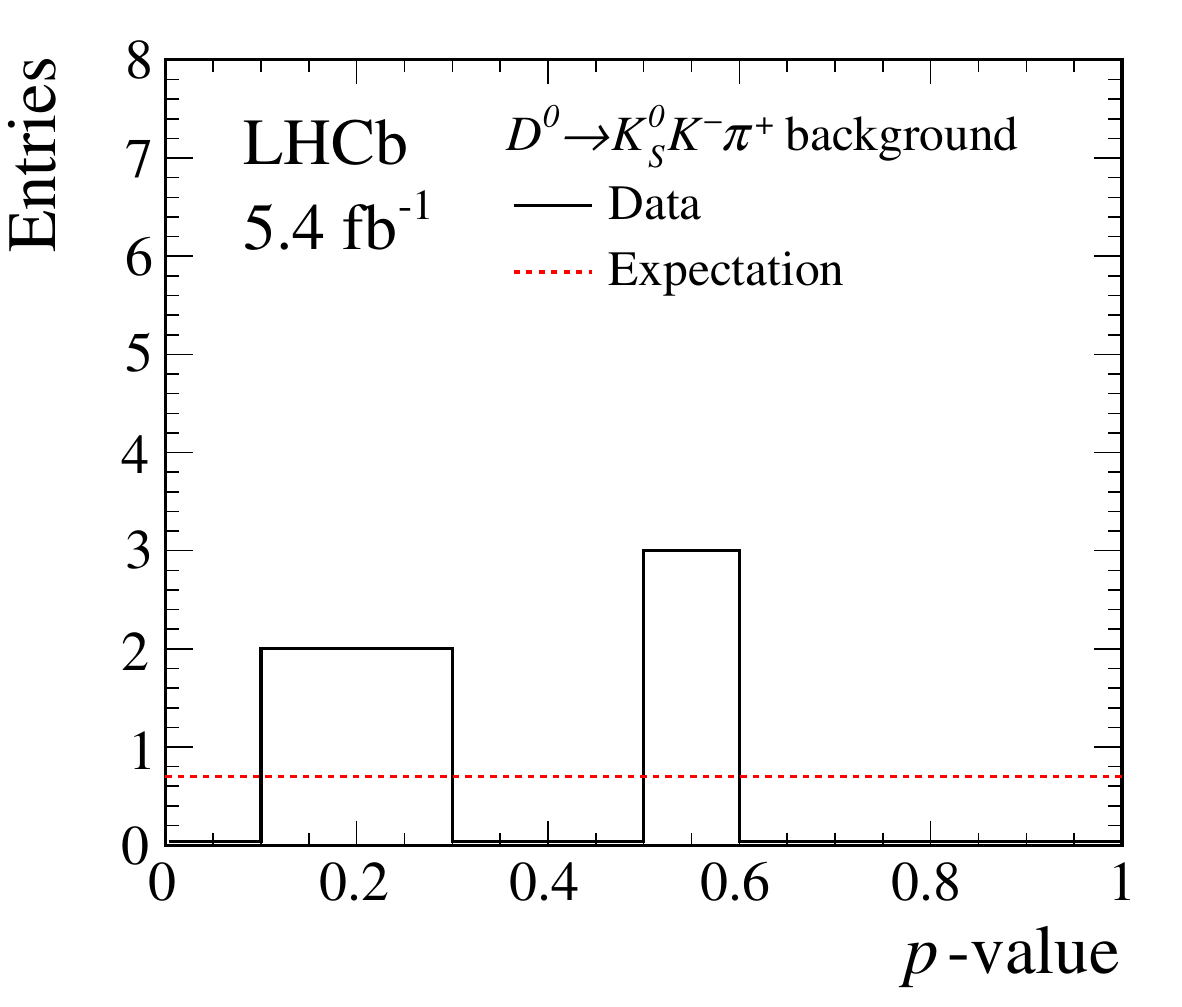}
    \includegraphics[width=0.45\linewidth]{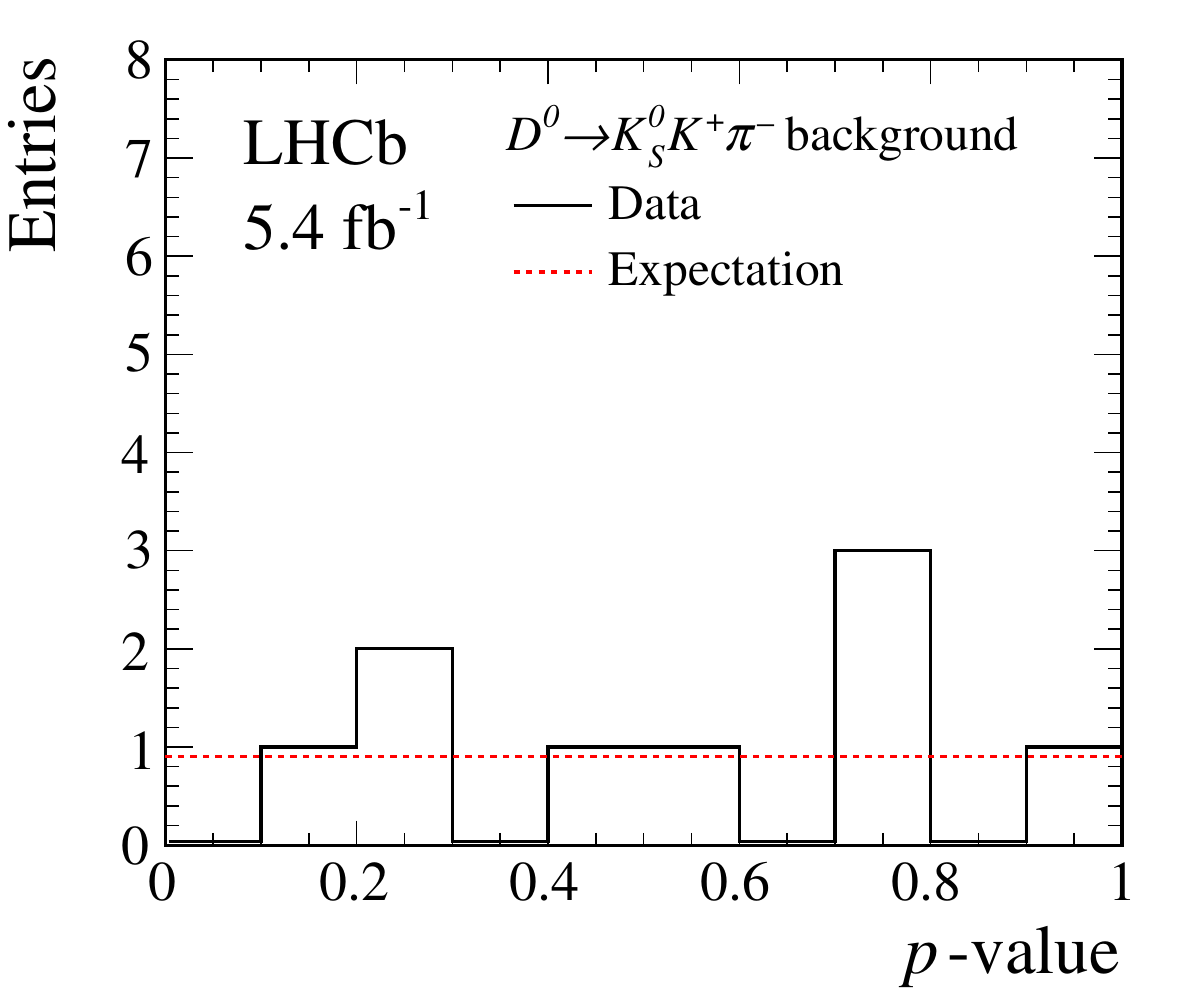}
    \caption{Distribution of $p$-values found in the (left) SS and (right) OS background-enhanced samples selected by considering a side-band in $\Delta M$. Each $p$-value is obtained by comparing datasets that contain roughly twice the background yield present in the signal window.}
    \label{fig:bkgpvalue}
\end{figure}

Background candidates are selected from the $\Delta M$ sidebands where \mbox{$|\Delta M- 145.4| > 1.2\mevcc$}, while otherwise requiring the standard signal selection. This sideband corresponds to the value of $\Delta M$ lying more than five times the detector resolution from the known value for the signal modes.
This dataset typically contains genuine \Dz and \KS mesons, whereas the \Dstarp candidate is formed by combining the \Dz meson with a random pion. This dataset is split into subsamples that contain twice the number of background events found in the signal region, determined using the fits described in Sect.~\ref{sec:selection}. 
The energy test is run on each of these `background-enhanced' subsamples. Again, this results in distributions of $p$-values, as shown in Fig.~\ref{fig:bkgpvalue}. These distributions are compared to the expected flat distribution (assuming no significant asymmetry in the background processes) using $\chi^2$-tests. The $p$-values of these tests are 3\% (for the SS mode) and 33\% (for the OS mode). While the SS mode yields only a percent-level compatibility with no underlying background effect, it is worth noting that no individual test in this case returns a $p$-value below 10\%, \emph{i.e.} at a level that would be indicative of any underlying background asymmetries. In addition, the seven $p$-values for the SS mode are combined and an overall test is performed by calculating a single overall statistic $P = -2 \sum_{i=1}^{N} \log p_i$, following Fisher's method as also discussed in Ref.~\cite{LHCb-PAPER-2023-005}. The $p$-value for this new statistic is found to be 25\%. Consequently, there is no significant indication that the presence of background causes bias in the searches for \CP violation presented here.

Nevertheless, an additional test is performed, considering candidate decays where the \KS candidate mass, the \Dz candidate mass, and the value of $\Delta M$ are all larger than the known signal values by more than 5 standard deviations of the detector resolution in each variable, while also requiring that  $m(\KS) < 562 \mevcc$, $m(\Dz) < 1966 \mevcc$ and $\Delta M < 165 \mevcc$.  Applying this selection provides roughly the same number of events in the signal region for both the SS and OS samples, so the energy test is run once on these samples, yielding $p$-values of 75\% and 17\%, again consistent with a \CP symmetric background. These large $p$-values further confirm that the presence of background in the signal channel does not mimic the effects of \CP violation.

Additionally, possible sources of background within the signal selection window were investigated. Samples of $\Dz \to \KS\pip\pim\pi^0$, $\Dz \to \KS\Kp\Km$
 and $\Dz \to \KS\pip\pim$
decays were simulated. These decays can mimic the signal process when the neutral pion is missed and/or when one of the charged particles is misidentified. The selection requirements applied in the analysis reduce these backgrounds to per mille level constructions and smaller, such that no significant asymmetry is expected in the signal channels from these sources. In addition, these background decays are expected to also be selected in the background-enhanced samples discussed above, where no evidence is found for the presence of background that can mimic \CP violation.

\section{Results and conclusions}
\begin{figure}[H]
    \centering
    \includegraphics[width=0.45\linewidth]{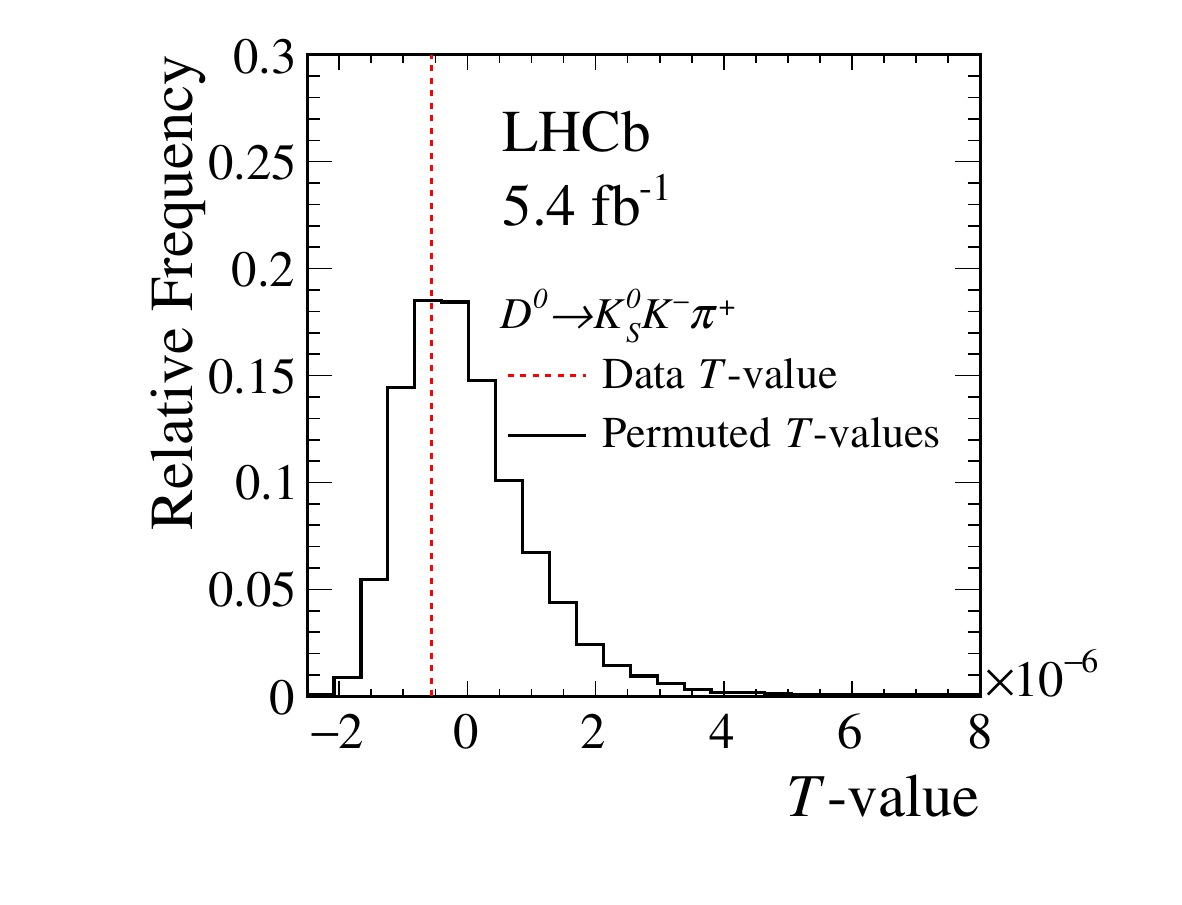}
    \includegraphics[width=0.45\linewidth]{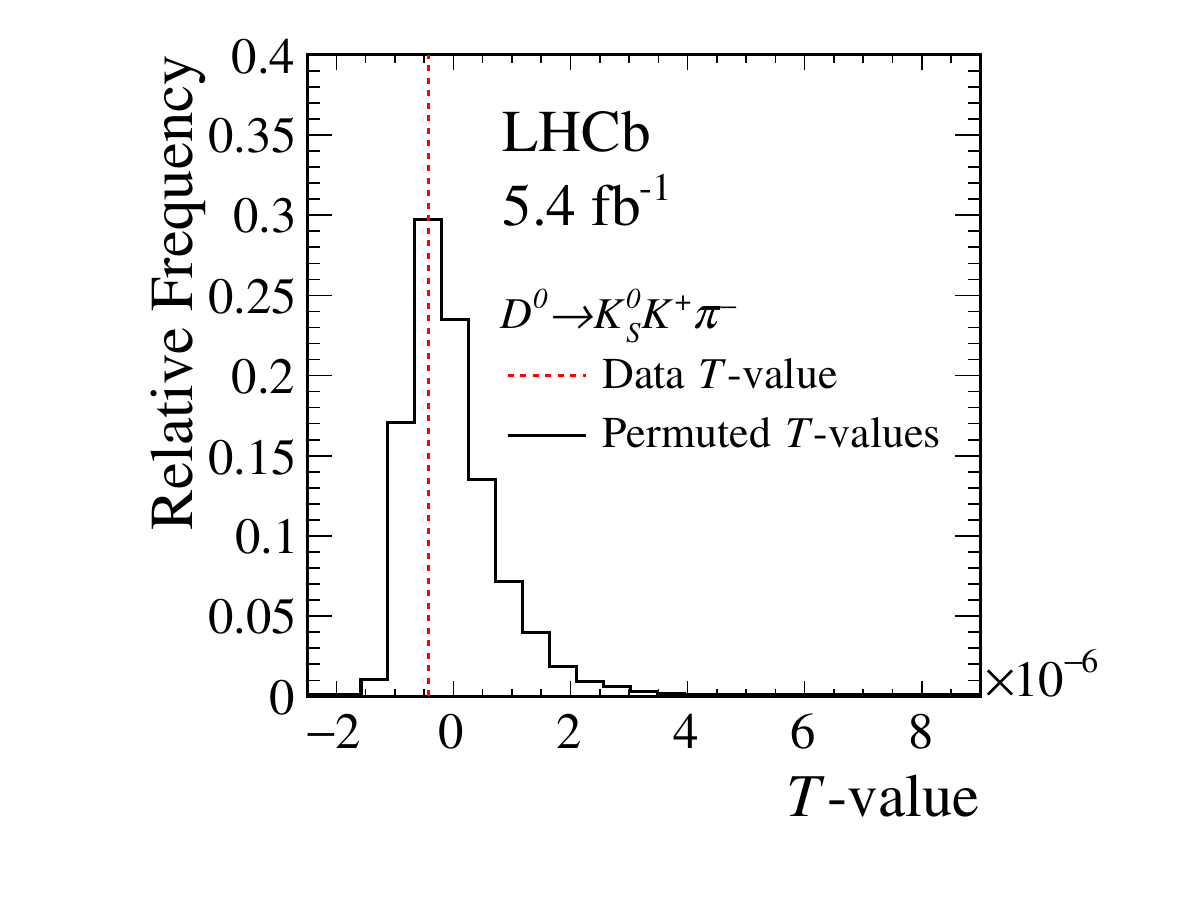}
    \caption{Value of $T$ found in data for (left) SS and (right) OS decays shown as a dotted red line. The normalised distribution of $T$-values expected under the hypothesis of local \CP symmetry is superimposed.}
    \label{fig:finalpvalue}
\end{figure}

Following confirmation that the method is free from any significant bias, either due to instrumentation effects or the presence of background, the signal regions are analysed using the energy test. This analysis considers the largest sample of \mbox{$\Dz \to \KS \Kpm \pimp$} decays studied to date, collected at the LHCb experiment in the second period of LHC operations, with yields of about $950$ thousand and about $620$ thousand candidates for the SS and OS sample, respectively, which are more than seven times larger than those considered in previous analyses of the same decay modes using data collected at the LHCb experiment~\cite{LHCb-PAPER-2015-026}. Figure~\ref{fig:finalpvalue} shows the $T$-value observed in data for each decay mode, alongside the distribution of $T$-values expected under the hypothesis of local \CP symmetry. 
The SS sample returns a $p$-value for the hypothesis of local \CP symmetry of 70\%, while the OS sample returns a $p$-value for this hypothesis of 66\%. 
Consequently, no evidence is found for local \CP violation in the phase space of \mbox{$\Dz \to \KS \Kpm \pimp$} decays. 

\section*{Acknowledgements}
%
%
\noindent We express our gratitude to our colleagues in the CERN
accelerator departments for the excellent performance of the LHC. We
thank the technical and administrative staff at the LHCb
institutes.
We acknowledge support from CERN and from the national agencies:
CAPES, CNPq, FAPERJ and FINEP (Brazil); 
MOST and NSFC (China); 
CNRS/IN2P3 (France); 
BMBF, DFG and MPG (Germany); 
INFN (Italy); 
NWO (Netherlands); 
MNiSW and NCN (Poland); 
MCID/IFA (Romania); 
MICINN (Spain); 
SNSF and SER (Switzerland); 
NASU (Ukraine); 
STFC (United Kingdom); 
DOE NP and NSF (USA).
We acknowledge the computing resources that are provided by CERN, IN2P3
(France), KIT and DESY (Germany), INFN (Italy), SURF (Netherlands),
PIC (Spain), GridPP (United Kingdom), 
CSCS (Switzerland), IFIN-HH (Romania), CBPF (Brazil),
Polish WLCG  (Poland) and NERSC (USA).
We are indebted to the communities behind the multiple open-source
software packages on which we depend.
Individual groups or members have received support from
ARC and ARDC (Australia);
Key Research Program of Frontier Sciences of CAS, CAS PIFI, CAS CCEPP, 
Fundamental Research Funds for the Central Universities, 
and Sci. \& Tech. Program of Guangzhou (China);
Minciencias (Colombia);
EPLANET, Marie Sk\l{}odowska-Curie Actions, ERC and NextGenerationEU (European Union);
A*MIDEX, ANR, IPhU and Labex P2IO, and R\'{e}gion Auvergne-Rh\^{o}ne-Alpes (France);
AvH Foundation (Germany);
ICSC (Italy); 
GVA, XuntaGal, GENCAT, Inditex, InTalent and Prog.~Atracci\'on Talento, CM (Spain);
SRC (Sweden);
the Leverhulme Trust, the Royal Society
 and UKRI (United Kingdom).

\addcontentsline{toc}{section}{References}
\bibliographystyle{LHCb}
\bibliography{main,standard,LHCb-PAPER,LHCb-CONF,LHCb-DP,LHCb-TDR}

\newpage
\centerline
{\large\bf LHCb collaboration}
\begin
{flushleft}
\small
R.~Aaij$^{35}$\lhcborcid{0000-0003-0533-1952},
A.S.W.~Abdelmotteleb$^{54}$\lhcborcid{0000-0001-7905-0542},
C.~Abellan~Beteta$^{48}$,
F.~Abudin{\'e}n$^{54}$\lhcborcid{0000-0002-6737-3528},
T.~Ackernley$^{58}$\lhcborcid{0000-0002-5951-3498},
B.~Adeva$^{44}$\lhcborcid{0000-0001-9756-3712},
M.~Adinolfi$^{52}$\lhcborcid{0000-0002-1326-1264},
P.~Adlarson$^{78}$\lhcborcid{0000-0001-6280-3851},
H.~Afsharnia$^{11}$,
C.~Agapopoulou$^{46}$\lhcborcid{0000-0002-2368-0147}
% ...

{\footnotesize \it

% ...
$^{11}$Universit{\'e} Clermont Auvergne, CNRS/IN2P3, LPC, Clermont-Ferrand, France\\
% ...
$^{35}$Nikhef National Institute for Subatomic Physics, Amsterdam, Netherlands\\
% ...
$^{44}$Instituto Galego de F{\'\i}sica de Altas Enerx{\'\i}as (IGFAE), Universidade de Santiago de Compostela, Santiago de Compostela, Spain\\
% ...
$^{46}$European Organization for Nuclear Research (CERN), Geneva, Switzerland\\
% ...
$^{48}$Physik-Institut, Universit{\"a}t Z{\"u}rich, Z{\"u}rich, Switzerland\\
% ...
$^{52}$H.H. Wills Physics Laboratory, University of Bristol, Bristol, United Kingdom\\
% ...
$^{54}$Department of Physics, University of Warwick, Coventry, United Kingdom\\
% ...
$^{57}$School of Physics and Astronomy, University of Glasgow, Glasgow, United Kingdom\\
$^{58}$Oliver Lodge Laboratory, University of Liverpool, Liverpool, United Kingdom\\
% ...
$^{78}$Department of Physics and Astronomy, Uppsala University, Uppsala, Sweden, associated to $^{57}$\\
% ...
}
\end{flushleft}

\end{document}